\documentclass[12pt,preprint]{aastex}

\shorttitle{{\it Spitzer} Observations of Nearby M Dwarfs}
\shortauthors{Riaz, Mullan & Gizis}

\begin{document}

\title{{\it Spitzer} Observations of Nearby M Dwarfs}

\author{Basmah Riaz, D. J. Mullan, John E. Gizis}
\affil{Department of Physics and Astronomy, University of Delaware,
    Newark, DE 19716; basmah@udel.edu, mullan@udel.edu, gizis@udel.edu}

\begin{abstract}

We present {\it Spitzer} IRAC and MIPS observations for a sample of eight M dwarfs: six dMe, one dM, and one sdMe star. All of our targets are found to have SEDs which are fitted within the error bars by a purely photospheric spectrum out to 24$\micron$. We find no evidence for IR excess. None of our targets is detected in the MIPS 70 and 160$\micron$ bands. The estimated ages for all are $>$10 Myr, suggesting that enough disk dissipation has occurred within the inner several AU of the star. For four of these, Mullan et al. (1989) had reported IRAS detections at 12$\micron$, although the reported fluxes were below the 5-$\sigma$ IRAS detection limit ($\sim$0.2 Jy). Mullan et al. also pointed out that V-K colors in dMe stars are larger than those in dM stars, possibly because of the presence of a chromosphere. Here we suggest that metallicity effects provide a better explanation of the V-K data. 

For reasons of observational selection, our targets are {\it not} the most active flare stars known, but being dMe stars indicates the presence of a chromosphere. Scaling from Houdebine's model of the AU Mic chromosphere, we have computed the free-free infrared excesses for a range of densities. Our {\it Spitzer} 24$\micron$ data shows that the chromospheres in two of our targets are less dense than in AU Mic by a factor of 10 or more. This is consistent with the fact that our sample includes the less active flare stars. Our models also indicate that the chromospheric contribution to the observed AU Mic emission at submillimeter wavelengths is only about 2\%.

\end{abstract}


\section{Introduction}

The collapse of the cold molecular cloud, from which stars are born, requires the presence of rotationally supported disks in order to conserve the angular momentum. On a timescale of $\sim$ $10^5$ yrs, a central protostar is formed surrounded by an infalling envelope and an accreting disk (Hartmann 2005). These young primordial disks are massive and have an initial gas:dust ratio similar to the interstellar ratio. As the system evolves, the primordial disk gradually dissipates, resulting in a decrease in the total disk mass on a timescale of $\sim$ $10^7$ yrs. The dissipation occurs through the sticking collisions of the smaller particles to form larger bodies, or planetesimals (e.g., Hillenbrand 2005). The shattering collisions among the planetesimals produce dust grains that form the secondary $\arcsec$debris$\arcsec$ disks. A debris disk was discovered around Vega (Aumann et al. 1984) by the Infrared Astronomical Satellite (IRAS). This was followed by various studies on the IRAS observations of debris disks around low-mass stars (e.g., Waters et al. 1987, Tsikoudi 1988, Tsikoudi 1989, Mullan et al. 1989, Iyengar 1991, Mathioudakis \& Doyle 1993). However, some of the recent follow-up observations could not confirm the IR excess emission seen for the IRAS detected low-mass stars (e. g., Song et al. 2002, Weinberger et al. 2004, Plavchan et al. 2005). Stars identified by IRAS were limited by its sensitivity. The IRAS Faint Source Catalog (FSC) has a 5-$\sigma$ detection limit of $\sim$0.2Jy at 12 and 25 $\micron$, and $\sim$0.28Jy at 60$\micron$, which is insufficient to detect debris disks around low-mass stars. At a distance of $\sim$50pc, IRAS could detect the photospheres at 12$\micron$ of only the stars with spectral types earlier than K0. At 25$\micron$, an M0 star would have to be within 5pc to be detected by IRAS.

The advent of {\it Spitzer Space Telescope} (Werner et al. 2004) has immensely expanded our knowledge of disk evolution. With sensitivities nearly 1000 times more than IRAS, and a wide wavelength range (3-160 $\micron$) offered by its instruments, {\it Spitzer} has made it possible to directly determine the IR emission from the circumstellar disks over a wide range of stellar masses and ages. 

Mullan et al. (1989; hereafter M89) conducted a search for dMe and dM stars in the IRAS database, and found a detection rate of 73\% and 82\% at 12$\micron$ for these two categories of stars, respectively. Based on the IRAS 12$\micron$ fluxes, these authors concluded that a dMe star will be 70\% brighter at this wavelength than a dM star, for the same K magnitude. Furthermore, 27\% of the dMe stars detected at 12$\micron$ were also detected at 100$\micron$. In order to confirm the excess IR emission detected by IRAS for dMe stars, we have carried out a {\it Spitzer} survey of a sample consisting of six dMe, one sdMe and one dM star, in the IRAC (Fazio et al. 2004) and MIPS (Rieke et al. 2004) bands. The sdMe and dM were included as control stars, that would allow comparisons within the sample. The stars in our sample are {\it not} the most active flare stars known. Because of previously approved {\it Spitzer} observing programs, our choice of targets was limited to somewhat less active stars. Despite this observational selection, our study is useful in the sense of allowing us to make comparisons with the results reported from IRAS data. In addition, if it turns out that we can detect IR excesses in our targets, or set limits on such excesses, this may provide a diagnostic for the presence of a chromosphere.

\section{Observations}

Observations in the IRAC and MIPS bands were obtained for our targets between October, 2004 and March, 2006. We could not obtain MIPS observations for GJ 83.1, as those were already scheduled for another program. For GJ 1284, we have MIPS channel 3 observations, while the ones in channels 1 and 2 have been obtained from the {\it Spitzer} archives. The IRAC observations were obtained in the subarray mode. We requested 1$\times$64 frames, with a frame time of 0.1s for each pointing. Additionally, a dither pattern of 4 positions Gaussian was requested, which allows the removal of instrumental artifacts, such as bad pixels. For MIPS observations, an exposure time of 3s was requested for channel 1, and 10s for channels 2 and 3. The number of cycles was 1, 2 and 4 for channels 1, 2 and 3, respectively.

Aperture photometry was performed on the artifact-corrected mosaic images using the task PHOT under the IRAF package APPHOT. We used an aperture radius of 3 pixels and a background annulus of 3-7 pixels for the IRAC data. The zero point fluxes of 280.9, 179.7, 115 and 64.1 Jy and aperture corrections of 1.124, 1.127, 1.143, and 1.234 were used for IRAC channels 1 through 4, respectively. The errors in magnitudes reported by PHOT are between $\sim$0.05 and 0.1mag in the IRAC bands. For the MIPS 24 $\micron$ data, an aperture radius of 6 pixels, background annulus of 12-17 pixels, zero point flux of 7.3 Jy and an aperture correction of 1.19 were used. The magnitude errors are between $\sim$0.05 and 0.2mag.  The calibration uncertainty in the IRAC and MIPS bands is $\sim$2\% and 10\%, respectively. The fluxes thus determined are listed in Table 1. None of our targets could be detected in the MIPS 70 and 160 $\micron$ bands.

\section{Discussion}

\subsection{Spectral Energy Distributions (SEDs)}

Fig. 1 shows the SEDs for our stars from the optical through the 24 $\micron$ bands. The optical magnitudes were obtained using the VizieR Service, and the JHK magnitudes from the 2MASS database. We have also included the IRAS fluxes for four of our targets that are present in the M89 sample. We have used the NextGen models (Hauschildt et al. 1999a,b) to fit the stellar photospheres. Gorlova et al. (2003) have determined the mean value for log g to be equal to 5.17 for field dwarfs. We have used the solar metallicity and log g=5.0 models for all of our stars. Table 2 lists the different parameters for our targets, along with the H$\alpha$ equivalent widths obtained from the PMSU online data (Hawley et al. 1996, Gizis et al. 2002). The distances were determined from the parallaxes (obtained from SIMBAD), and the effective temperatures were obtained from the spectral types using the relation in Kenyon \& Hartmann (1995). Mould (1976) has shown that the chief blanketing agents in M dwarfs are the molecules of $H_{2}O$ and TiO, but the I and J bandpasses are less affected by either of these bands than the V and R, or H, K and L. Thus we have normalized the models to the J-band, as it best represents the photospheric emission from the star.

As can be seen for all of the stars in our sample, the observed IRAC and MIPS points coincide with the photospheric model fits within the observational errors. None of our stars display excess emission in the IRAC and MIPS bands. The absence of any such excess raises questions about previously claimed high detection rate (73\%) at 12$\micron$ in IRAS data. We discuss this point further in section 3.2. 

As we have already noted in the Introduction, the stars in our sample are not members of the group of {\it most active} dMe stars. As a result, IR excesses which may rise due to free-free processes are expected to be more difficult to detect among our sample stars than among the most active flare stars. The observed lack of IR excesses will be used below (section 3.4) to set upper limits on chromospheric density.

We can rule out ages of 10 Myr or younger for our targets, since their red spectra (from PMSU surveys) do not show signatures of low surface gravity. These stars are close to the main sequence, and therefore unlikely to be T Tauri-type stars. The mean inner disk ($<$ 0.1AU) lifetime from various disk surveys in the near-IR wavelengths is found to be around 2-3 Myr, with no evidence for near-IR excess found beyond 5 Myr (Hillenbrand 2005). Haisch et al. (2001) conducted a L-band survey of clusters at different ages, and found that the timescale for all the stars to lose their disks in the inner $\sim$10AU is about 6 Myr. Strong stellar winds or planet formation can explain the rapid dissipation of dust within the inner several AU of the star (e.g., Plavchan et al. 2005, Weinberger et al. 2004, Liu et al. 2004). The ages for our stars suggest that enough disk dissipation has occurred to leave little or no signs of disks around them. However, cooler remnant disks might be detectable at submillimeter wavelengths, similar to the one found around the dMe star AU Mic. This young ($\sim$12Myr) active flare star shows no excess in the IRAC bands (Chen et al. 2005). However, excess emission detected in the MIPS 70$\micron$ band, and at sub-mm wavelengths indicate the presence of a debris disk at a temperature of $\sim$40K (Liu et al. 2004, Chen et al. 2005). Chen et al. have discussed that the micron-sized grains around AU Mic have lifetimes much shorter than the stellar lifetimes, but much longer than the inferred grain lifetime under stellar wind drag, suggesting that collisions between larger bodies must be the source of grain destruction. Emission from such a debris disk might make detection of chromospheric emission more difficult in AU Mic, but older stars (including the ones in our sample) present more favorable conditions for extracting information about the chromosphere.

\subsection{Detection Limits}

A non-detection at 70 and 160 $\micron$ suggests that our exposures were not deep enough. Table 3 lists the predicted photospheric fluxes at these wavelengths, as determined from the NextGen model fits to the observed data. Using the set up we had requested for observations in these two bands, SENS-PET returned us with medium-background point source sensitivities of 2710 and 18,000 $\mu$Jy (1-$\sigma$) in the 70 and 160 $\micron$ bands, respectively. This suggests that while the photospheres of none of our targets could have been detected at 160$\micron$, only three have photospheric fluxes higher than the detection limit (1-$\sigma$) at 70$\micron$. However, there is a high probability that a 1-$\sigma$ detection is caused by noise, as compared to a 5-$\sigma$ detection. A 5-$\sigma$ detection limit would be 13.55mJy and 90mJy at 70 and 160 $\micron$, respectively. This is higher than the predicted photospheric fluxes for any of the stars, which explains why none of them could be detected at these longer wavelengths. However, an excess emission as strong as $\sim$10 and $\sim$100 times more than the 5-$\sigma$ detection limit at 70 and 160 $\micron$, respectively, could have been detected with the set-up requested.

At 12$\micron$, IRAS Faint Source Catalog (FSC) has a 5-$\sigma$ detection limit of $\sim$0.2 Jy. Typical noise levels at this wavelength are 0.05 Jy. Out of the 55 probable identifications at 12$\micron$ in M89 sample of dMe stars, 22 have fluxes below this limit. This drops the 73\% detection rate reported by these authors to 44\%, and suggests that the detections in their sample with IRAS 12$\micron$ fluxes less than $\sim$0.2 Jy may be false. Song et al. (2002) could not confirm the IR excess seen for three tentative stars, identified by correlating IRAS FSC and Hipparcos catalog, with their ground based observations.  These authors have concluded that the detection threshold bias (or the Malmquist bias) is responsible for the bogus IRAS IR excesses. This bias is explained as caused by the occasional large upward noise fluctuations, which could boost the IR fluxes, and result in the star being classified as an IR-excess source. This can significantly affect the low S$\slash$N sources, rather than the high S$\slash$N ones.

Four of our targets are present in the M89 survey; GJ 83.1, GJ 207.1, GJ 398 and GJ 781. Their reported observed fluxes at 12$\micron$ are 0.14$\pm$0.033Jy, 0.13$\pm$0.032Jy, 0$\pm$0.046Jy (zero flux means the peak flux is below $\sim$2-$\sigma$ limit), and 0.06$\pm$0.026Jy, respectively. The uncertainties are 1-$\sigma$ values. In Fig. 1, these points are in excess for GJ 207.1 and GJ 781, though no excess is seen at 24$\micron$. Comparing with the predicted photospheric fluxes at 12$\micron$ (Table 3) shows that these are not significant excesses. Thus for these four common targets, it is not surprising that an excess emission is also not seen in the {\it Spitzer} observations. M89 have argued that such sources (with fluxes $<$ 0.1 Jy at 12$\micron$) cannot be spurious IRAS sources, since the distribution of $R_{oc}$ (the ratio of observed to calculated fluxes) is symmetric about unity. For spurious sources, this ratio would be very large. Thus the IRAS measurements may not be completely unreliable.

At 100$\micron$, M89 report a detection rate of 27\% for stars that were detected at 12$\micron$. One of our targets (GJ 781) is among these stars, with a flux at 100$\micron$ of 2.15$\pm$0.36 Jy. The predicted photospheric flux is 0.0003 Jy, and explains the strong excess seen at this wavelength in the SED of this star. However, we could not detect it at 70 and 160 $\micron$, which rules out the IRAS detection. As discussed in the IRAS Explanatory Supplement, at 100$\micron$, the infrared sky is characterized by emission from interstellar dust, known as $\arcsec$infrared cirrus$\arcsec$. This can generate either a well-defined point source or an extended source. On the basis of this, Mathioudakis \& Doyle (1993) have suggested that unless a source detected at 100$\micron$ was also detected at the shorter wavelengths, and the flux at 12$\micron$ is higher than that at 100$\micron$, the detection at 100$\micron$ should be treated with caution. The ratio of the 12 to 100$\micron$ fluxes is $\sim$0.03 for GJ 781, which is very low, and explains its non-detection in the MIPS bands.

\subsection{Metallicity Effects}

M89 found that at the same spectral subclass, the mean value of the V-K color for a dMe star is larger than that for a dM star. This is evident from Fig. 2, where the offset seen for the dMe in their sample from the mean dM relation (from Hawley et al. 1996), is as large as $\sim$2 for the coolest stars. The V-K and TiO5 values were obtained from the PMSU3 online data (Hawley et al. 1996, Gizis et al. 2002), while the absolute K magnitudes were derived using the K magnitudes from 2MASS, and the distances determined from parallaxes. The errors in TiO5 indices are estimated to be $\pm$0.02-0.04. M89 concluded that the excess emission detected at 12$\micron$ by IRAS explains the larger V-K colors for the dMe stars. We suggest that metallicity effects provide a better explanation for the V-K data. Bonfils et al. (2005) have shown that for the same absolute $M_{K}$ magnitude, metal-rich M dwarfs are redder in V-K than the metal-poor ones. These authors have defined a polynomial relation between $M_{K}$ and V-K, that can be used to determine [Fe$\slash$H]:

\begin{equation}
[Fe/H] = 0.196 - 1.527 ~M_{K} + 0.091 ~M_{K}^2 + 1.886 ~(V-K) - 0.142~ (V-K)^2,
\end{equation}

\noindent valid for $M_{K}$ between 4 and 7.5, (V-K) between 2.5 and 6, and [Fe/H] between -1.5 and +0.2, with a dispersion of 0.2 dex. In Fig. 3, we have used their isometallicity contours for [Fe$\slash$H] that vary from -1.1 dex to +0.9 dex, in steps of 0.1 dex. We note that their relations are valid for [Fe$\slash$H] between -1.5 and +0.2 dex, and so we are not certain about the metallicities of stars that fall out of these limits. The spread in the V-K color, for a given TiO5 index, seen for the dMe stars in M89 sample can be explained by their metallicity differences. Although the stars in our sample do not show a large offset in their V-K colors, the small offsets become significant when metallicity effects are taken into account. Four of these have near-solar metallicities, while the other four are metal-poor. The one sdMe star in our sample (GJ 781) is the most metal poor ([Fe$\slash$H] $\sim$ -0.9), as expected. Bonfils et al. (2005) have discussed that two mechanisms work together to decrease the luminosity of the more metal-rich stars through the visible filters, thus resulting in redder V-K colors for the higher metallicity stars. One is that for a given mass, higher metallicity decreases the bolometric luminosity. The other effect is that an increase in metallicity results in a shift in the flux from the visible to the near-IR, due to line-blanketing by the TiO and VO bands.

\subsection{Search for IR Signatures of a Chromosphere/Corona}

Based on the presence of strong emission in the Balmer lines and in other visible and UV lines, it is well known that a dMe star possesses a chromosphere where the temperatures rise to values well in excess of the photospheric temperature. Modelling of such chromospheres in terms of non-LTE radiative transfer can be quite sophisticated (e.g., Houdebine 1990). In the Sun, there is also a pronounced temperature rise above the photospheric value in the chromosphere and corona. Compared to the solar atmosphere, the densities in active dMe star chromospheres can be larger than the solar values (at a given temperature) by factors of 10 or more (e.g., Cram and Mullan 1979).  

Now in the Sun, as a result of free-free processes, the chromospheric temperature rise maps into an excess of solar brightness temperature $T_{b}$ at wavelengths longer than roughly 200$\micron$ (e.g., Gu et al. 1997).  In a star where the chromospheric density exceeds the solar value, the onset of an IR excess is expected to occur at wavelengths less than 200$\micron$. In view of this, we now examine the {\it Spitzer} data in the context of setting a limit on chromospheric densities in our targets.

Our starting point is the model chromosphere which was derived by Houdebine (1990) for one of the most active flare stars known, AU Mic.  Using emission line profiles of H$\alpha$ and H$\beta$, Houdebine derived a semi-empirical model in which the temperature is assumed to increase linearly with log m (where {\it m} is the mass loading, in units of g $cm^{-2}$ ) from a temperature minimum of 3255 K up to T = 8000 K at log m =  -2.9. Above T = 8000 K, the temperature rises sharply to T = 3 $\times$ $10^{5}$ K. We consider a number of models in which the electron densities are assumed to be uniformly larger or smaller than the Houdebine model by factors of 0.01-10. 

For each model, for a line of sight coming into the star from outside, we compute the free-free optical depth for each level in the atmosphere for a given wavelength using the absorption coefficient given by Kraus (1966):  

\begin{equation}
                  K~ (cm^{-1})  = 1.2 \times 10^{-30} ~N_{e}^{2}~ \lambda_{\mu}^2 ~\slash ~T^{1.5},
\end{equation}

\noindent where $N_{e}$ is electron density ($cm^{-3}$),  $\lambda_{\mu}$ is wavelength in microns, and T is temperature (K).  At long enough wavelengths, the chromosphere has an optical depth that is so large we do not ÔseeÕ all the way in to the photosphere. For such wavelengths, once the free-free optical depth is tabulated for a particular wavelength, the emergent flux at that wavelength can be computed using a standard result for flux integrals in radiative transfer (Aller 1963):

\begin{equation}
                    F(\tau = 0)  =   0.8839 ~B(\tau = 0.397) + 0.1161~ B(\tau = 2.723), 
\end{equation}

\noindent where B($\tau$) is the Planck function evaluated at the specified optical depth. At all wavelengths of interest to us here, we are on the Rayleigh-Jeans tail of the continuum flux from a dMe star. Assuming that infra-red emission emerges uniformly from the disk of the star, the emergent flux can therefore be converted simply to a brightness temperature $T_{b}(\lambda)$ at wavelength $\lambda$ in terms of the temperatures in the model at two particular values of optical depth: 

\begin{equation}
                     T_{b}(\lambda)  = 0.8839 ~T(\tau = 0.397) + 0.1161 ~T(\tau = 2.723).
\end{equation}

\noindent We have used this formula to compute the brightness temperature expected at each wavelength. For short wavelengths, the chromosphere is optically thin, and $T_{b}(\lambda)$ reduces to the photospheric temperature.

Results for models with various densities are shown as a function of wavelength in Fig. 4. The piecewise nature of the curves is due to the piecewise nature of HoudebineÕs table.
The denser the chromosphere, the shorter the wavelength at which the chromospheric rise in temperature becomes apparent in the $T_{b}$ curve. For the model labeled den=10, with a density which is 10 times larger than the AU Mic model, the chromosphere causes $T_{b}$ to begin to increase at wavelengths as short as 1$\micron$. For a model with den = 0.1, one must go to wavelengths of 100$\micron$ before $T_{b}$ begins to increase above the photospheric value. (Thus, based on the results of Gu et al. (1997), den=0.1 may be considered as more or less representative of the solar atmosphere.)
 
We have converted the observed fluxes at various wavelengths from two target stars into $T_{b}$ for purposes of comparison. The limiting behavior of $T_{b}$ is to become constant, independent of wavelength, if the photospheric flux dominates at a given wavelength. This occurs at short wavelengths. 

For the two stars with {\it Spitzer} data plotted in Fig. 4, we see that even at $\lambda$ = 24$\micron$, the observed brightness temperatures do not show any significant increase above the photospheric value. This result is not consistent with either of those two stars having a chromosphere with den=1 or 10. That is, if either star had a chromosphere that is as dense as that in AU Mic, or 10 times denser, then the 24$\micron$ data from {\it Spitzer} would have shown a clear increase above the photospheric value. The {\it Spitzer} data are, however, compatible with models den=0.1 and den =0.01, i.e. models with densities which are less than the AU Mic chromosphere by factors of 10-100. 

This conclusion is consistent with the observational selection that is at work in our sample: we have chosen targets which are not the most active flare stars. To the extent that AU Mic is representative of the latter class of stars, the stars in our sample are expected in all cases to be less active than AU Mic. Therefore, it would not be surprising to find that the chromospheric densities in our sample stars fall below the AU Mic densities.

The theoretical curves in Fig. 4 suggest that we could better diagnose mean chromospheric densities in our target stars if we were able to obtain submillimeter data at sensitivities better than $\sim$0.2 mag. Depending on the sensitivity, and on the number of wavelength points, we could map the rise in temperature from photosphere into the chromosphere in dMe stars. This could provide valuable confirmation of the model structures which have been derived from optical line profiles (Houdebine 1990). 

However, in the case of AU Mic, where sub-mm data already exists (Liu et al. 2004, Chen et al. 2005), contamination from a debris disk at $\sim$40K overwhelms the chromospheric emission. Our models indicate that the chromospheric contribution to the observed AU Mic emission at sub-mm wavelengths is about 2\% (see Fig. 5). Even if we increase the density of the chromosphere to 10 times that of AU Mic, the chromospheric contribution is about 10\%. Thus the models estimate that almost none of the observed excess in AU Mic is due to free-free chromospheric emission.

\section{Summary}

We report results from a {\it Spitzer} IRAC and MIPS survey of six dMe, one dM and one sdMe star. The SEDs of all of our stars are purely photospheric; none display any excess in any of the IRAC or MIPS bands. We could not detect any of our stars in the 70 and 160$\micron$ bands. The absence of any IR excess emission could be explained by the fact that all of our stars have ages $>$ 10 Myr, suggesting that enough disk dissipation has occurred to leave little or no signs of disks around them. Four of our targets are present in M89's survey, but their fluxes are below the 5-$\sigma$ detection limit at this wavelength. At 100$\micron$, the observed IRAS flux for GJ 781 is 2.15 Jy, but the ratio of 12 to 100$\micron$ fluxes is too small, which suggests that the infrared cirrus could be responsible for the detection. We have also shown that metallicity effects can explain the larger V-K colors seen for dMe stars by M89, as compared to those of the dM stars, at the same spectral subclass. Although our sample does not consist of the most active flare stars known, these being dMe stars indicates the presence of a chromosphere. Scaling from Houdebine's model of the AU Mic chromosphere, we have computed the free-free infrared excesses for a range of densities. Our {\it Spitzer} data at 24$\micron$ indicates that the chromospheres in two of our targets are less dense than in AU Mic by a factor of 10 or more. This is consistent with the fact that our sample of stars includes the less active flare stars. Our models also indicate that the chromospheric contribution to the observed AU Mic emission at sub-mm wavelengths is only about 2\%.

\acknowledgments

This work is based on observations made with the {\it Spitzer Space Telescope}, which is operated by the Jet Propulsion Laboratory, California Institute of Technology under a contract with NASA. Support for this work was provided by NASA through an award issued by JPL/Caltech. This work has made use of the VizieR service, and the SIMBAD and 2MASS databases. We would like to thank the referee for his useful comments.

{\it Facilities:} \facility{{\it Spitzer Space Telescope}}

\clearpage

\begin{deluxetable}{cccccccc}
\tabletypesize{\scriptsize}
\tablecaption{IRAC and MIPS Observations}
\tablewidth{0pt}
\tablehead{
\colhead{Star} & \colhead{3.6 $\micron$} & \colhead{4.5 $\micron$}  &
\colhead{5.8 $\micron$} & \colhead{8 $\micron$}  & \colhead{24 $\micron$}  \\
\colhead{} & \colhead{mJy}  & \colhead{mJy}  & \colhead{mJy}  & \colhead{mJy} & \colhead{mJy}  
}
\startdata
GJ 1065	&	308.26	$\pm$	7	&	207.33	$\pm$	0.3	&	147.19	$\pm$	7.5	&	80.57	$\pm$	1.5	&	10.78	$\pm$	0.4	\\
GJ 1284	&	1028.96	$\pm$	146.6	&	658.06	$\pm$	110	&	456.19	$\pm$	77.8	&	239.42	$\pm$	68.4	&	31.74	$\pm$	8.4	\\
GJ 207.1	&	629.01	$\pm$	123	&	403.48	$\pm$	87.4	&	292.25	$\pm$	46.5	&	156.68	$\pm$	38.6	&	16.29	$\pm$	7.8	\\
GJ 398	&	321.56	$\pm$	52.5	&	219.21	$\pm$	27.1	&	148.80	$\pm$	18.4	&	81.30	$\pm$	16.1	&	10.78	$\pm$	0.7	\\
GJ 781	&	209.96	$\pm$	54.6	&	141.67	$\pm$	28.2	&	97.96	$\pm$	16.4	&	48.10	$\pm$	18.4	&	7.14	$\pm$	1	\\
GJ 83.1	&	852.69	$\pm$	82.4	&	575.81	$\pm$	40.8	&	406.30	$\pm$	19.2	&	225.59	$\pm$	23.3	&	--		\\
LHS 2320	&	103.07	$\pm$	5.4	&	70.48	$\pm$	0.2	&	48.00	$\pm$	1.7	&	26.70	$\pm$	1	&	3.05	$\pm$	0.3	\\
LP 467-16	&	207.43	$\pm$	10.7	&	141.85	$\pm$	0.4	&	99.14	$\pm$	3.8	&	53.50	$\pm$	3.1	&	7.69	$\pm$	1.1	\\
\enddata
\tablecomments{The 1-$\sigma$ errors include IRAC and MIPS calibration uncertainties of $\sim$2\% and 10\%, respectively.}
\end{deluxetable}

\clearpage

\begin{deluxetable}{ccccccccc}
\tabletypesize{\scriptsize}
\tablecaption{Different Quantities for Our Targets}
\tablewidth{0pt}
\tablehead{
\colhead{Star} & \colhead{Spec. Type} & \colhead{H$\alpha$ EW ($\AA$) (PMSU1)} & \colhead{H$\alpha$ EW ($\AA$) (PMSU3)} & \colhead{d (pc)} & \colhead{TiO5} & \colhead{$M_{K}$}& \colhead{V-K} & \colhead{[Fe$\slash$H]}  \\
}
\startdata
GJ 1065	&	M3.5	&	--	&	-0.3	&	9.5	&	0.53	&	6.0	&	4.6	&	-0.29	\\
GJ 1284	&	M3	&	3	&	2.8	&	10.8	&	0.33	&	8.4	&	5.6	&	0.02	\\
GJ 207.1	&	M2.5	&	8	&	4.6	&	15.1	&	0.42	&	6.9	&	5.0	&	-0.01	\\
GJ 398	&	M3.5	&	5.1	&	4.2	&	13.7	&	0.42	&	7.9	&	5.0	&	-0.12	\\
GJ 781	&	sdM1.5	&	1.9	&	--	&	16.7	&	0.47	&	6.2	&	4.8	&	-0.85	\\
GJ 83.1	&	M4.5	&	1.7	&	--	&	4.5	&	0.30	&	7.3	&	5.5	&	-0.09	\\
LHS 2320	&	M5	&	14.5	&	--	&	21.7	&	0.28	&	8.6	&	6.2	&	-0.03	\\
LP 467-16	&	M5	&	8.6	&	--	&	8.5	&	0.77	&	7.0	&	3.9	&	0.02	\\
\enddata

\end{deluxetable}

\clearpage

\begin{deluxetable}{cccc}
\tabletypesize{\scriptsize}
\tablecaption{Predicted Photospheric Fluxes at 12, 70 and 160 $\micron$}
\tablewidth{0pt}
\tablehead{
\colhead{Star} & \colhead{12 $\micron$} & \colhead{70 $\micron$} & \colhead{160 $\micron$} \\ 
\colhead{} & \colhead{mJy} & \colhead{mJy} & \colhead{mJy}  \\
}
\startdata
GJ 1065	&	42.63	&	1.23	&	0.20	\\
GJ 1284	&	154.31	&	4.51	&	0.71	\\
GJ 207.1	&	97.53	&	2.82	&	0.45	\\
GJ 398	&	48.30	&	1.48	&	0.23	\\
GJ 781	&	34.30	&	0.96	&	0.15	\\
GJ 83.1	&	124.15	&	3.89	&	0.61	\\
LHS 2320	&	15.34	&	0.46	&	0.07	\\
LP 467-16	&	30.20	&	0.90	&	0.14	\\
\enddata

\end{deluxetable}

\clearpage
\begin{figure}
 \begin{center}
    \begin{tabular}{cc}
      \resizebox{60mm}{!}{\includegraphics[angle=270]{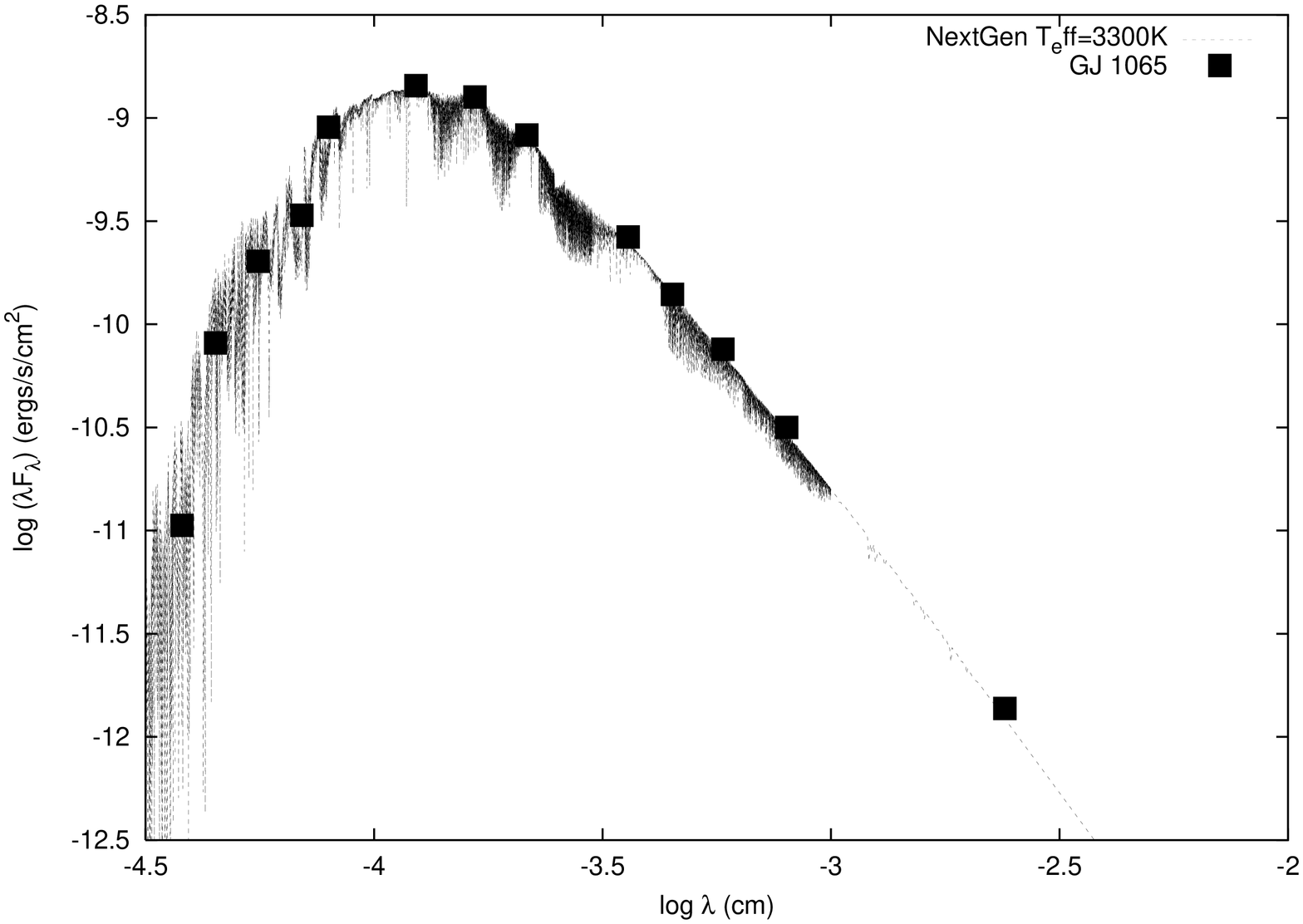}} &
      \resizebox{60mm}{!}{\includegraphics[angle=270]{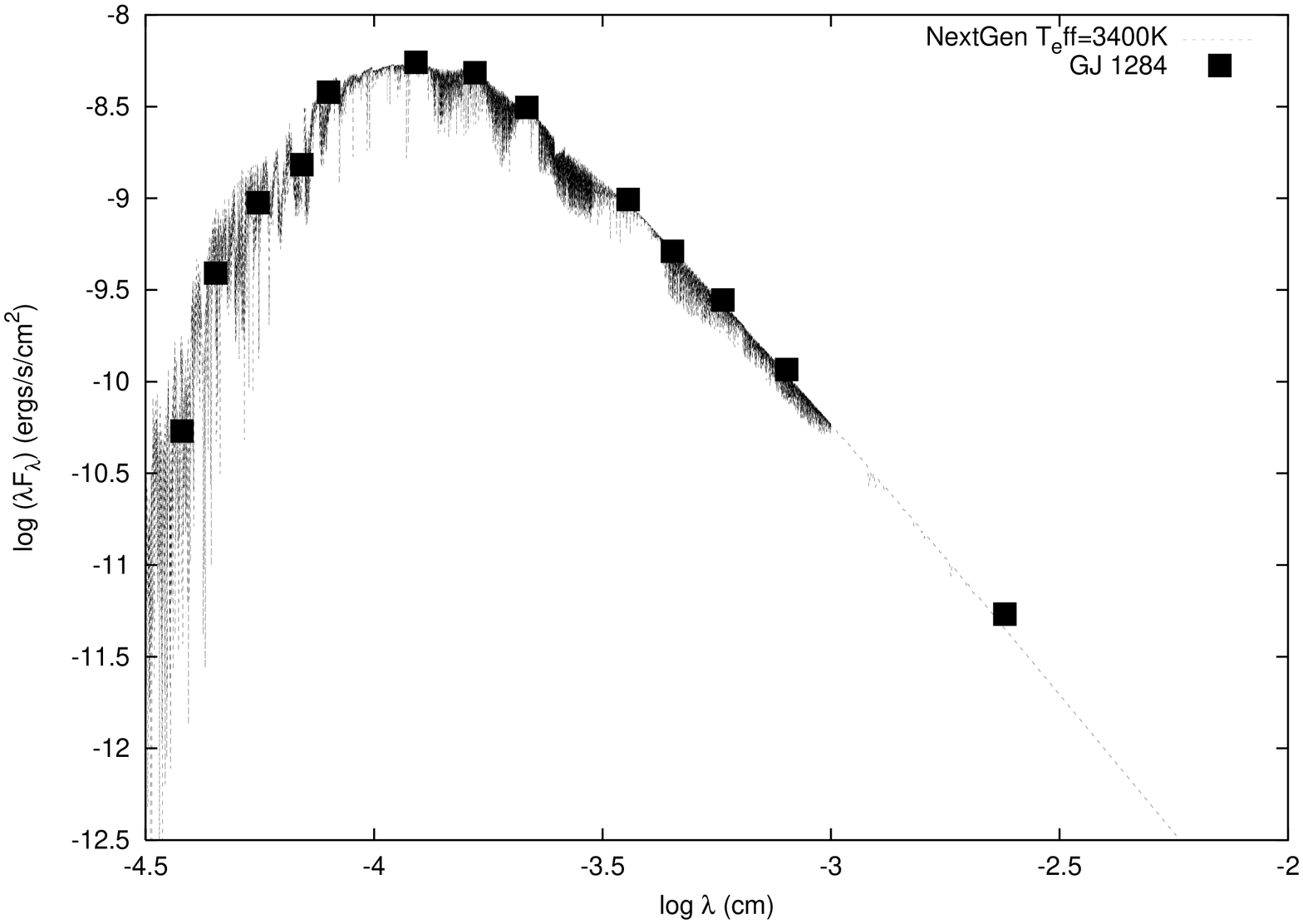}} \\
      \resizebox{60mm}{!}{\includegraphics[angle=270]{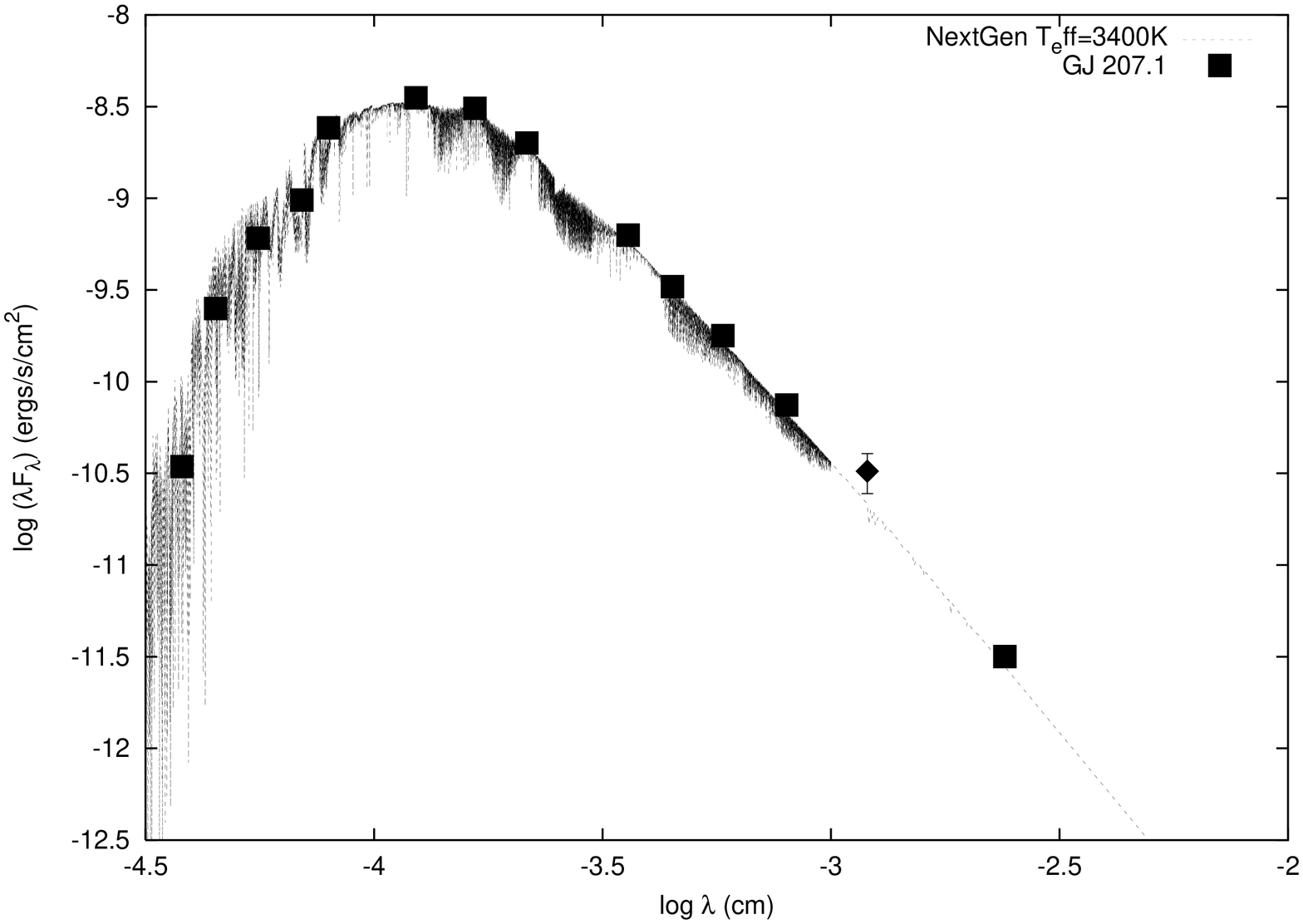}} &
      \resizebox{60mm}{!}{\includegraphics[angle=270]{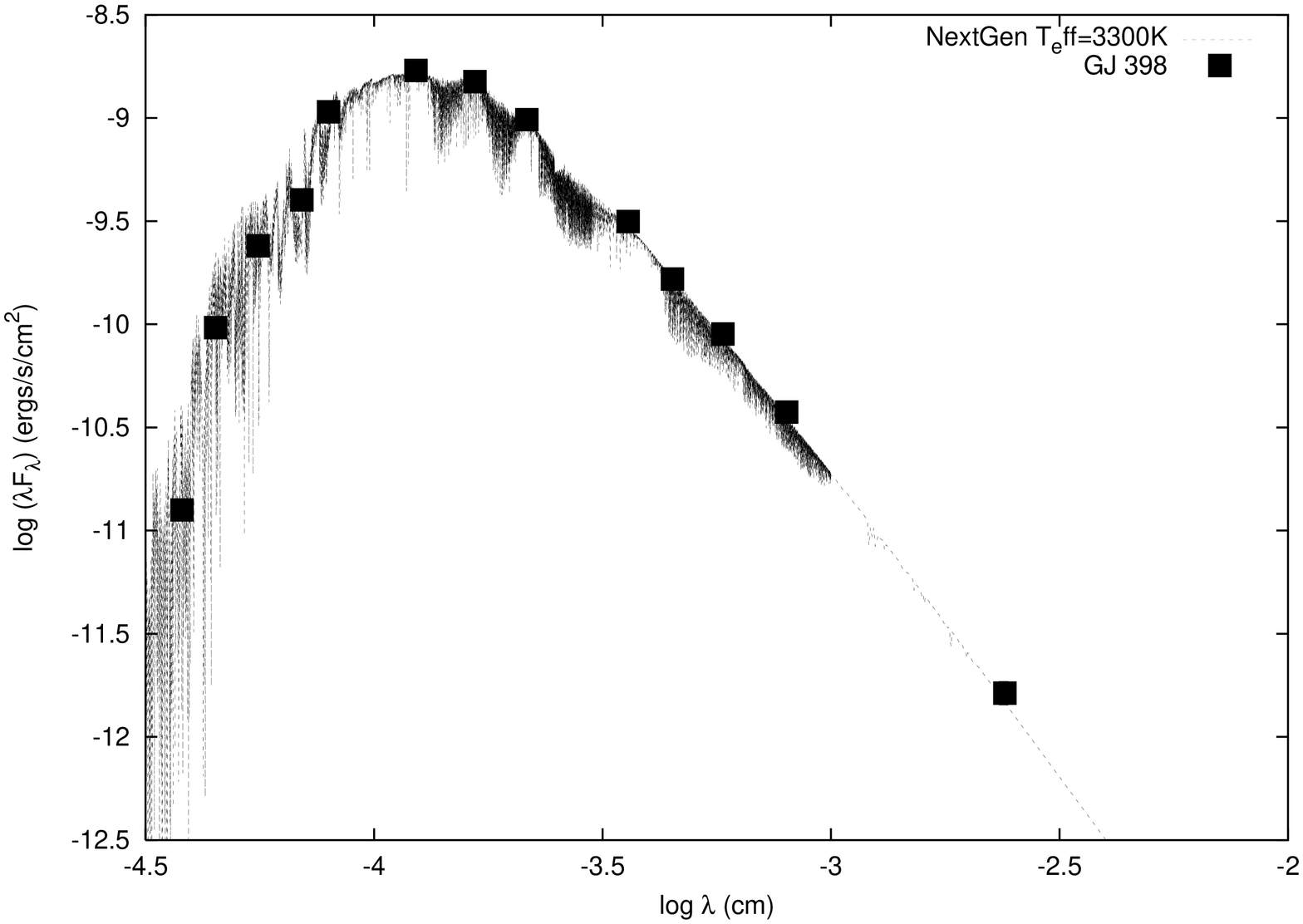}} \\
      \resizebox{60mm}{!}{\includegraphics[angle=270]{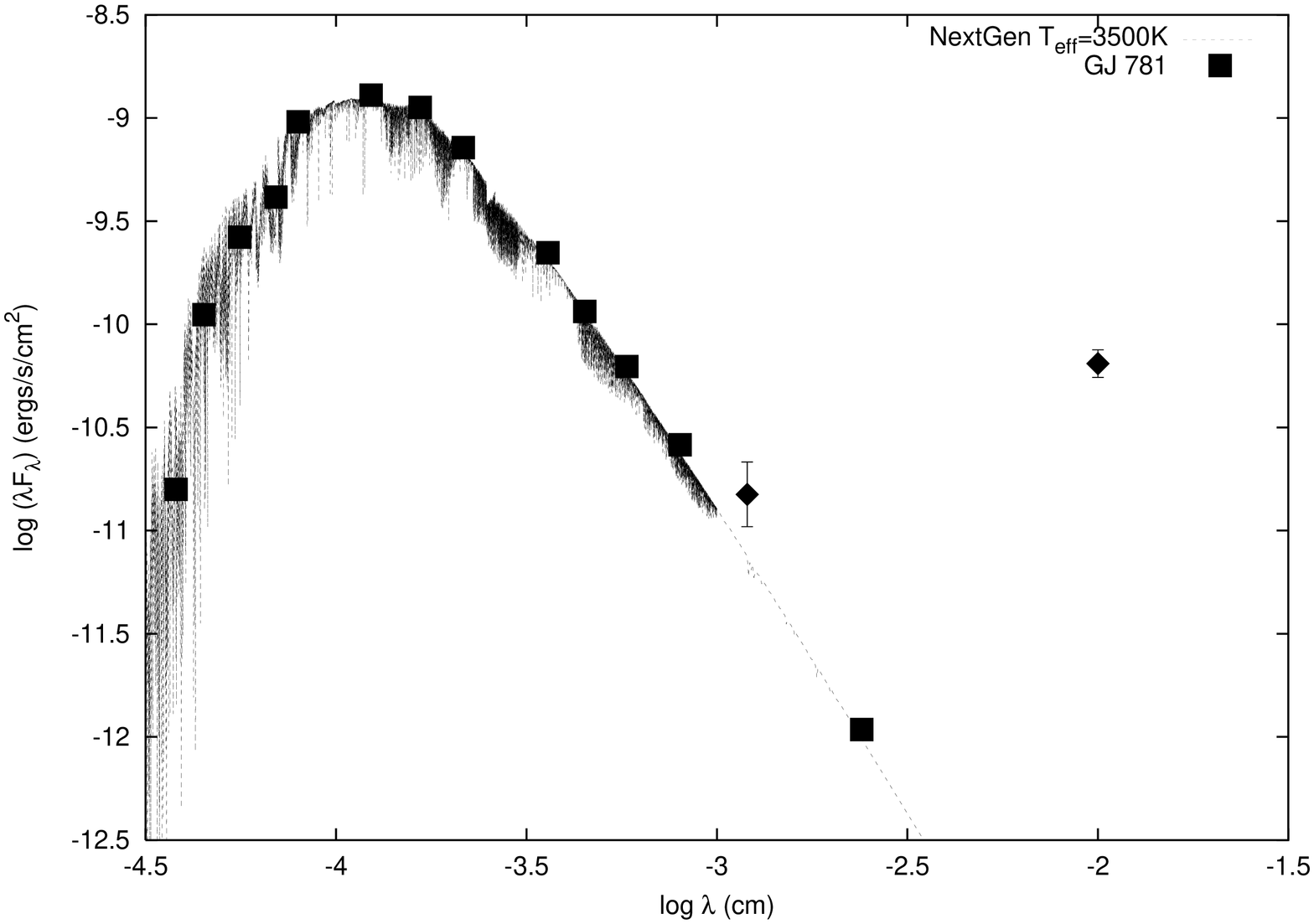}} &
      \resizebox{60mm}{!}{\includegraphics[angle=270]{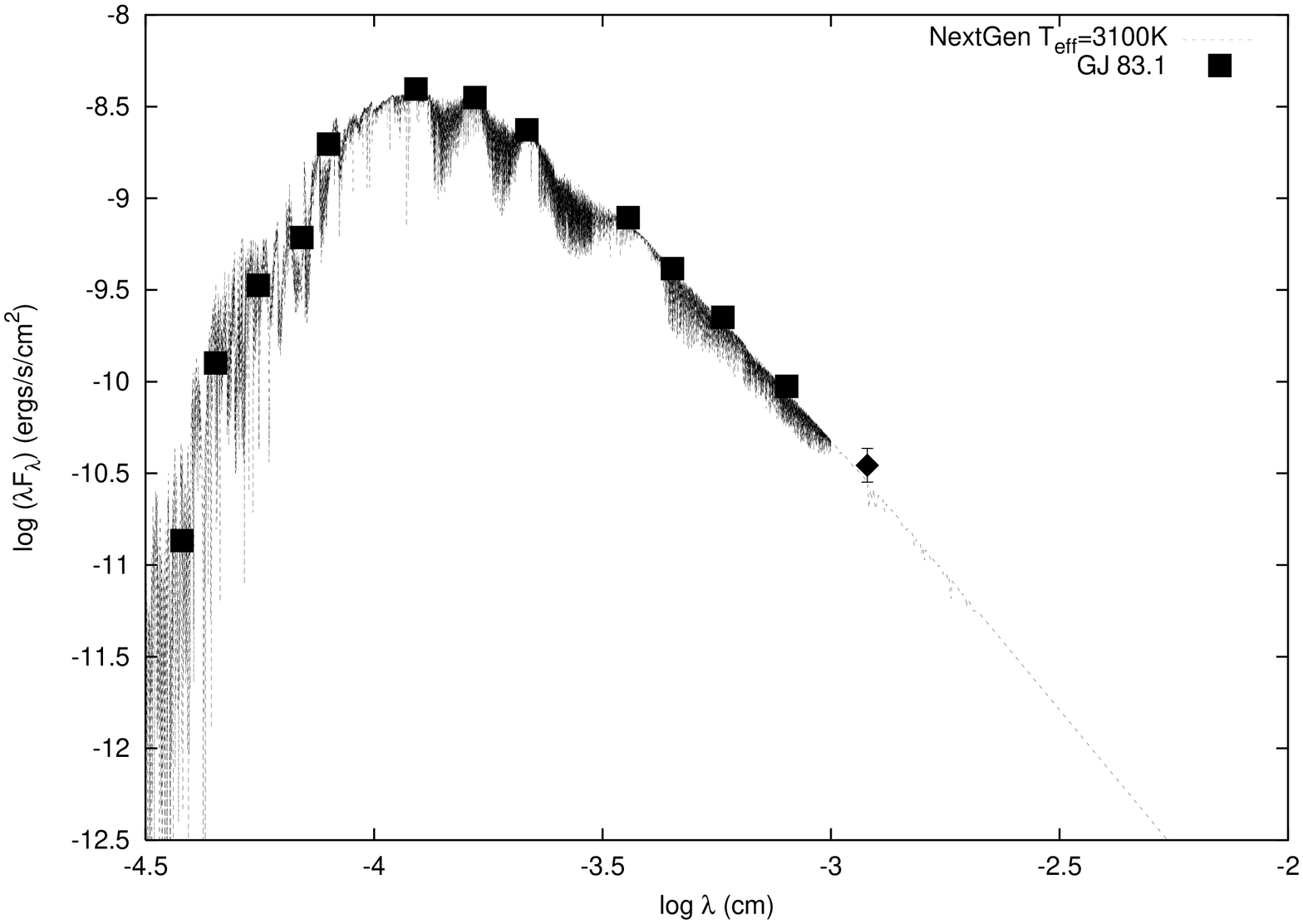}} \\
      \resizebox{60mm}{!}{\includegraphics[angle=270]{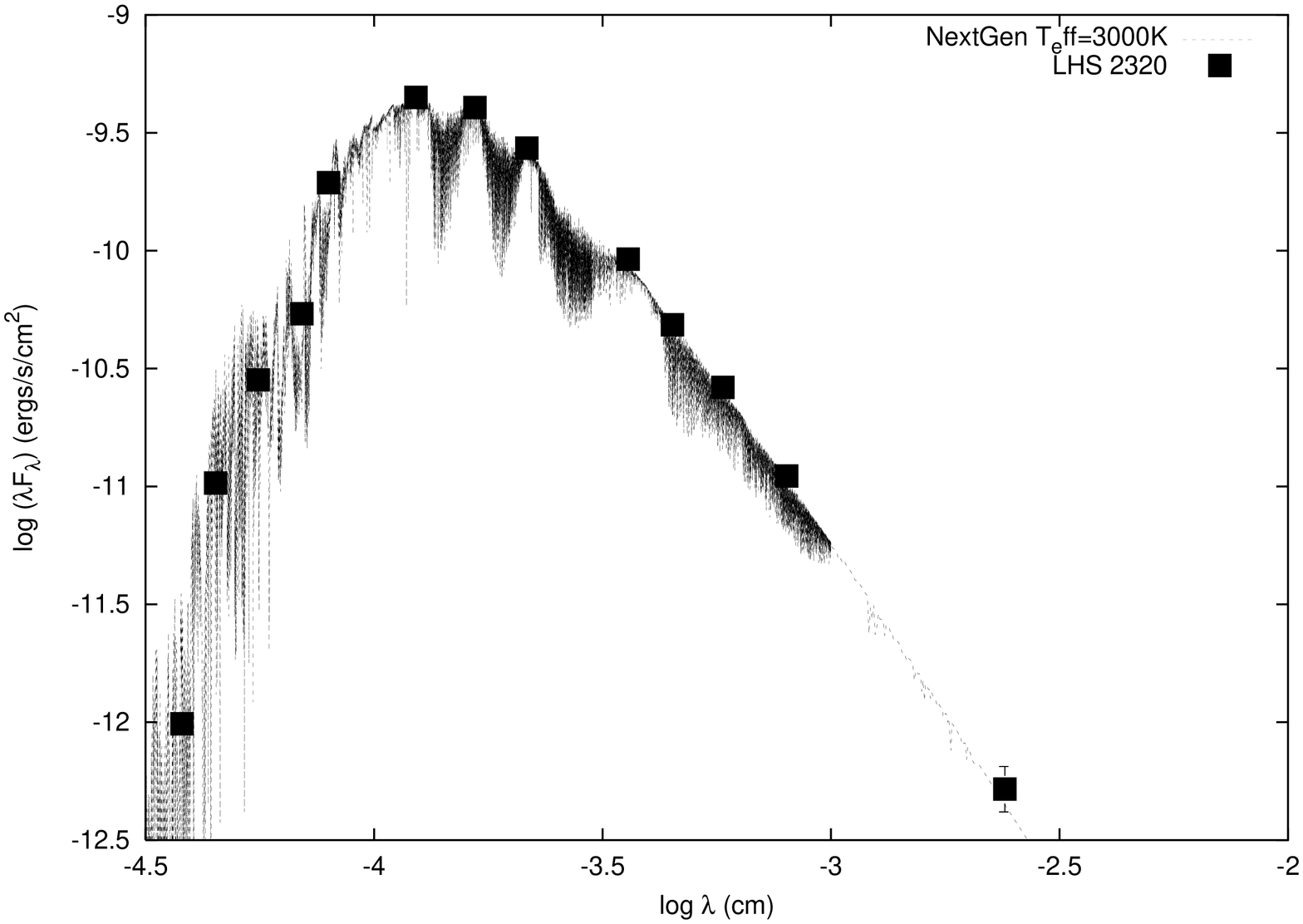}} &
      \resizebox{60mm}{!}{\includegraphics[angle=270]{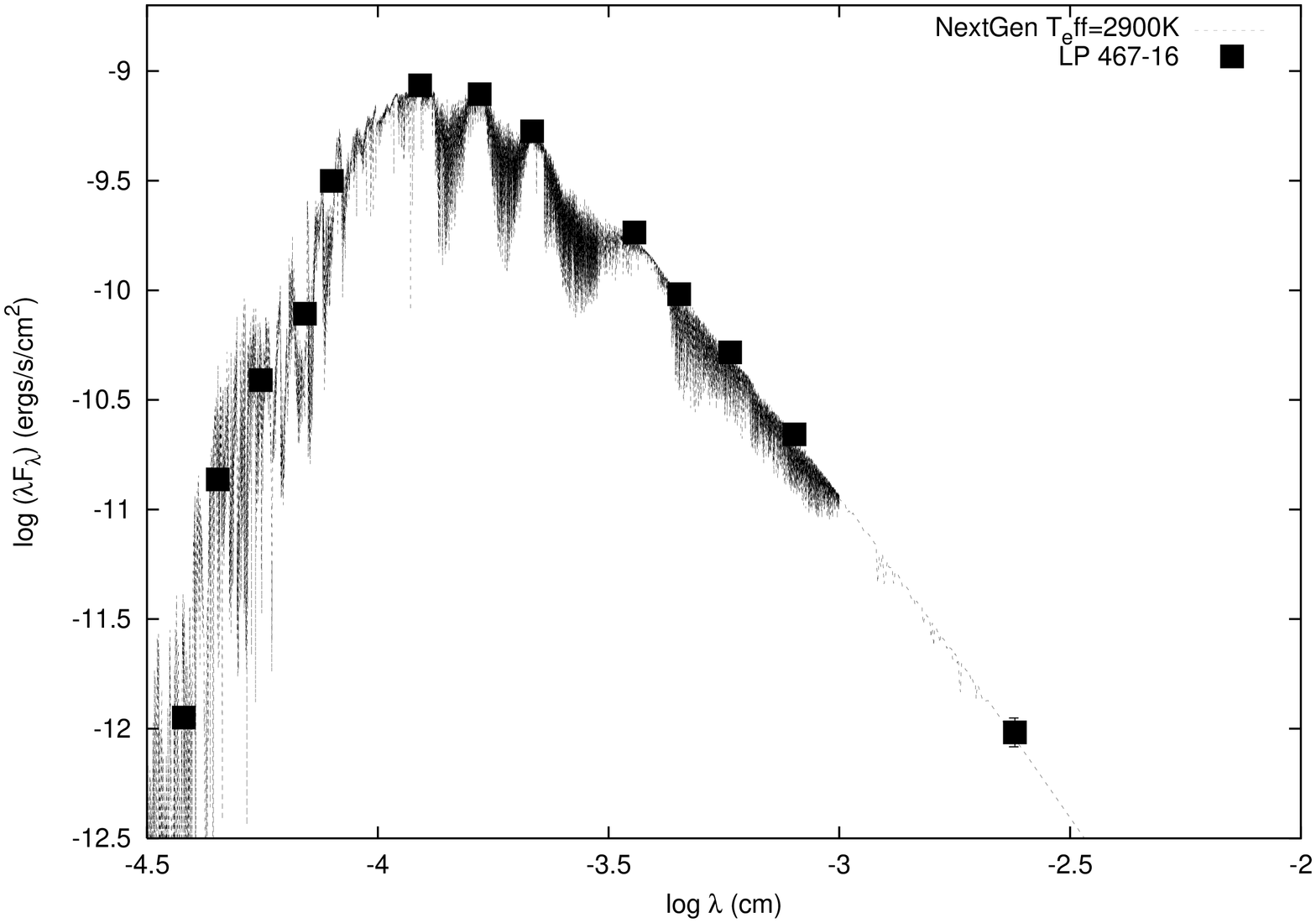}} \\
    \end{tabular}
    \caption{SEDs for the eight targets. Optical data from VizieR service, JHK from 2MASS, IRAC and MIPS from this work. Bold triangles represent IRAS data from M89. Dotted line represents the NextGen model. The 1-$\sigma$ error bars for the {\it Spitzer} data are smaller than the symbol size.}
  \end{center}
\end{figure}

\clearpage

\begin{figure}

\resizebox{150mm}{!}{\includegraphics[angle=270]{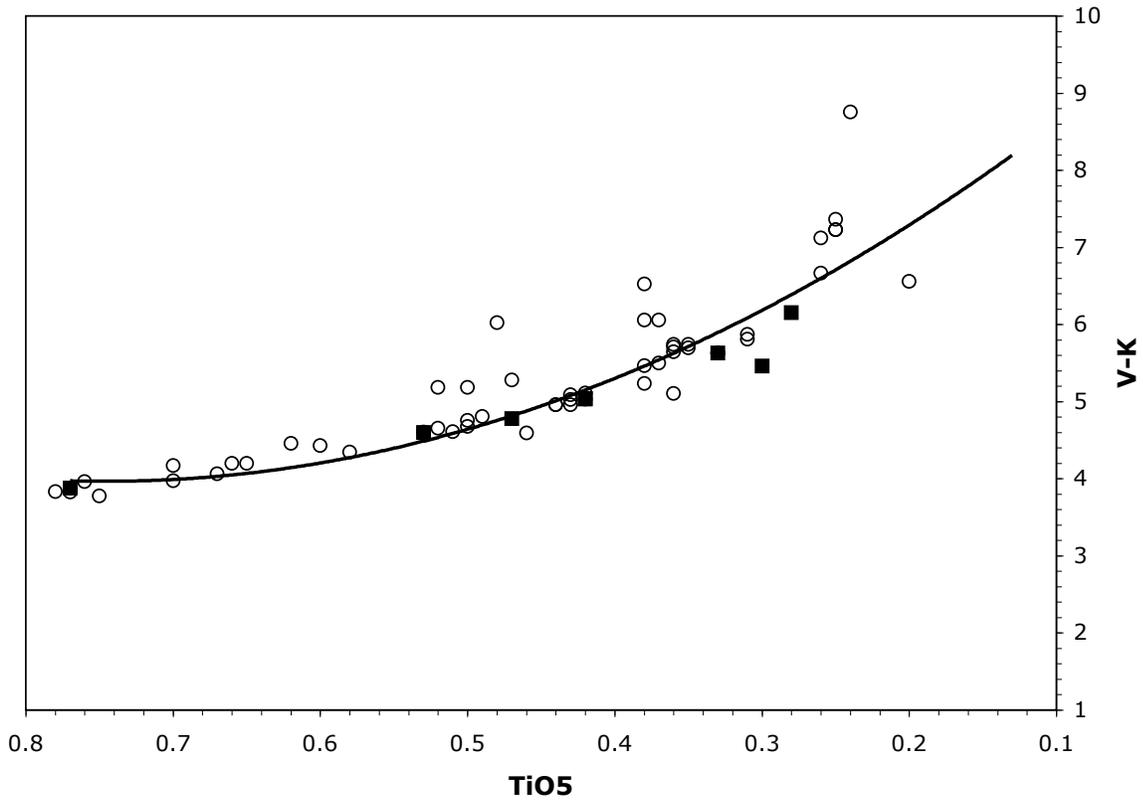}}
\caption{TiO5 indices vs. V-K colors. Filled squares -- this work, open circles -- stars from M89. Bold line represents the mean dM relation from Hawley et al. (1996).}
\end{figure}

\clearpage

\begin{figure}
\resizebox{150mm}{!}{\includegraphics[angle=270]{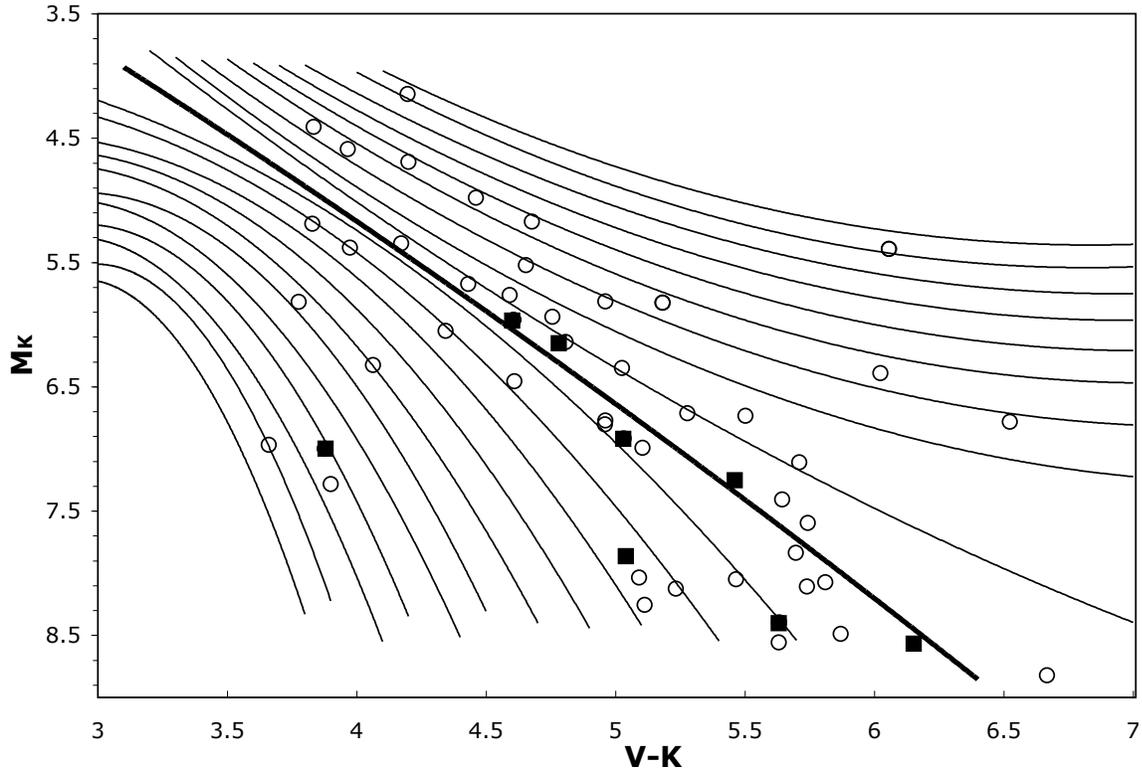}}
\caption{$M_{K}$ vs. V-K colors. Symbols same as in Fig. 2. Thin solid lines represent the isometallicity contours from Bonfils et al. (2005), spaced by 0.1 dex from - 1.1 dex (bottom) to +0.9 dex (top). Thick solid line represents the solar metallicity contour.}
\end{figure}

\clearpage

\begin{figure}
\resizebox{150mm}{!}{\includegraphics[angle=270]{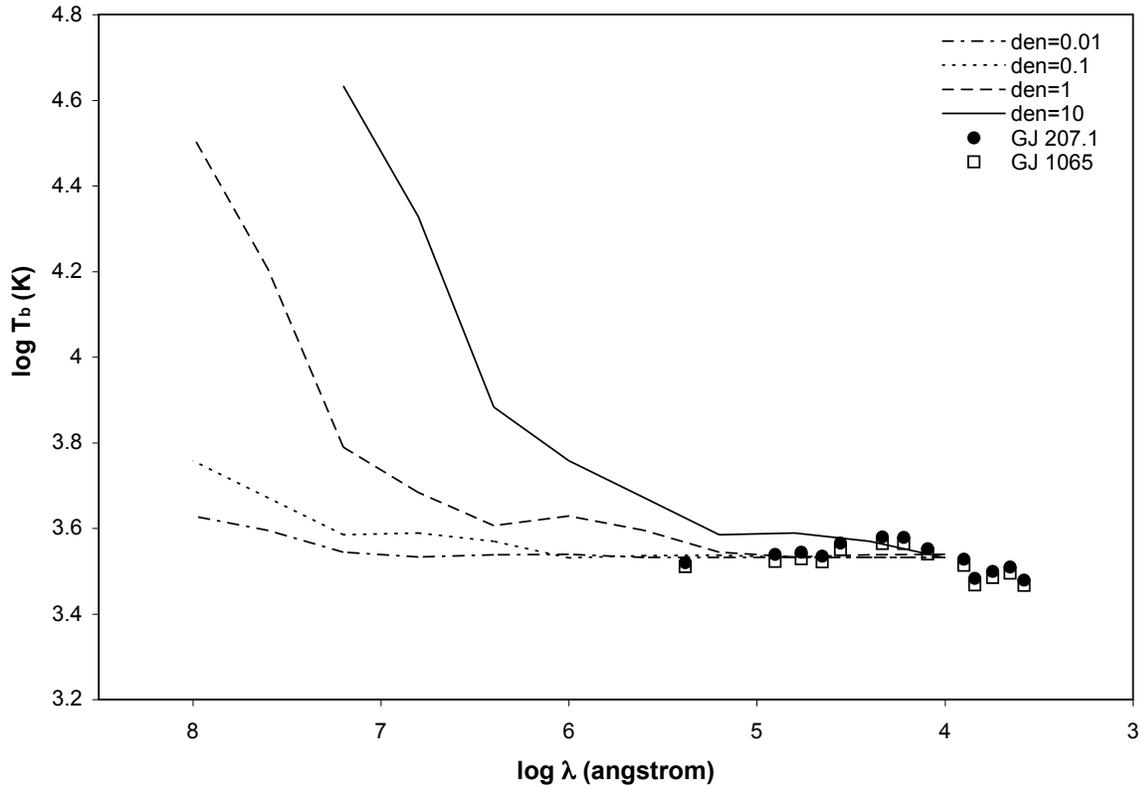}}
\caption{Model chromosphere for AU Mic by Houdebine (1990), for a range of electron densities between 0.01 to 10 times that of AU Mic. Also shown are observed data points for GJ 207.1 and GJ 1065. The 1-$\sigma$ errors are between 0.001 and 0.02, and are smaller than the symbol size.}
\end{figure}

\clearpage

\begin{figure}
\resizebox{150mm}{!}{\includegraphics[angle=270]{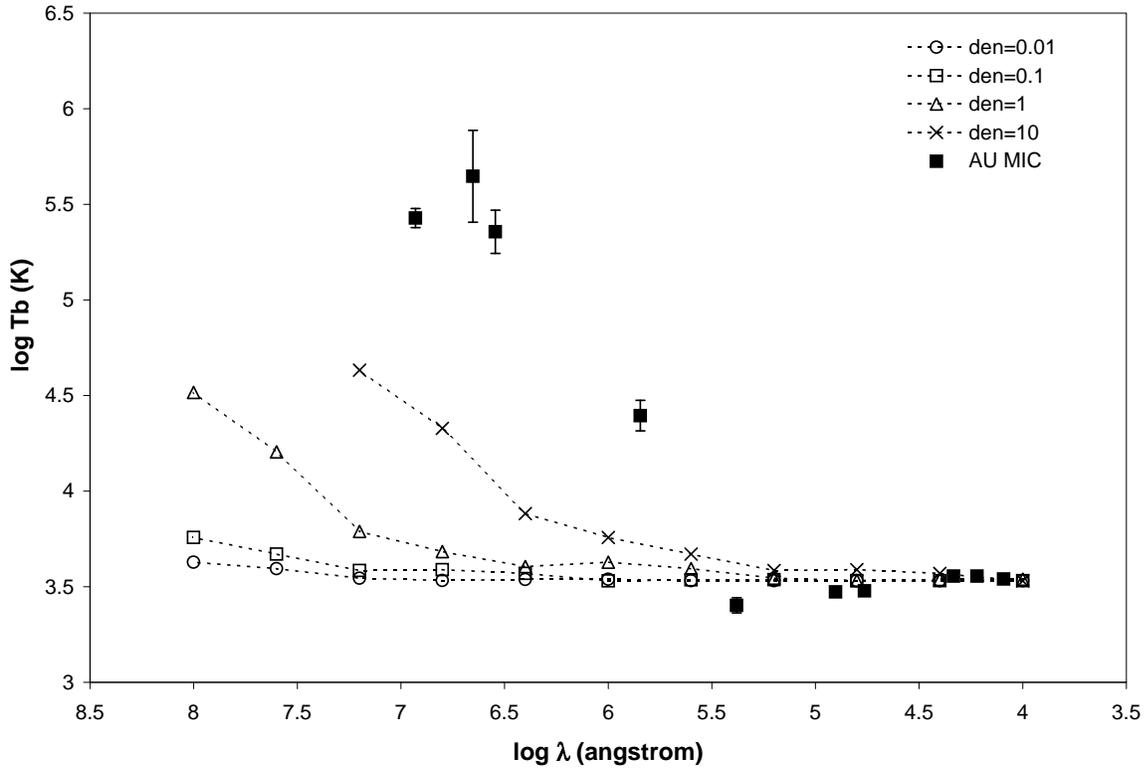}}
\caption{Models same as in Fig. 4. Bold squares represent observed IRAC, MIPS and submillimeter data for AU Mic from Chen et al. (2005) and Liu et al. (2004).}
\end{figure}

\end{document}